\documentstyle[twoside,12pt]{memsait}

\begin{opening}
\title{THE CONTRIBUTION OF GALAXIES TO $\Omega_0$}
\author{Paolo SALUCCI$^1$ \&  Massimo PERSIC$^{1,2}$}
\institute{$^1$SISSA, Strada Costiera 11, 34014-Trieste, Italy}
\institute{$^2$Osservatorio Astronomico di Trieste, Trieste, Italy}
\date{}
\end{opening}
\begin{document}

\oddpagefooter{Proceedings "Second Italian Cosmology Meeting, Asiago"
}{}{\thepage}
\evenpagefooter{\thepage}{}{}
\begin{abstract}
We estimate the contribution of galaxies to the cosmological density $\Omega_0$
including also their extended dark halos. We find $\Omega_{gal}< 0.1$,
so implying that the matter in galaxies cannot by itself close the Universe.
\end{abstract}
\section{Introduction}
\par
In recent years important progress has been made in
successfully modelling the mass
structure of galaxies. These stem both from the acquisition of new and higher
quality observational data, and from a number of
 efforts aimed to devise theoretical
methods suitable to study this issue. For spiral galaxies, there are now
available several hundreds of rotation curves extending to the optical size
(see Mathewson et al. 1992) and about 50 (HI rotation curves) extending
 out to twice
their optical size (e.g., Broeils 1992). For elliptical galaxies, the
kinematics is well known inside $R_e$ (e.g., van der Marel 1991), but much less
known in the outer regions; however in several objects the presence of extended
rotating gaseous disks allows an alternative way of tracing the underlying
gravitational potential (see Bertola et al. 1993). Finally, the kinematics of
several tens of dwarf galaxies is obtained  from their rotation curves or
velocity dispersion (e.g., Pryor 1992).  It is well known that these
observations, in connection with the
observed  luminosity profiles, imply in galaxies a widespread presence  of
a non-luminous component (dark matter, hereafter DM) which turns out to be the
major contributor to the mass (see review by Ashman 1992).

Ellipticals and
spirals need very careful methods of mass modelling in order to take into
account the large radial variation of the dark-to-visible mass ratio within
individual galaxies as well as differences among different objects.
As a result, a simple picture for the mass structure of galaxies has emerged:
this is given by a very concentrated luminous component surrounded by a more
diffuse and extended non-luminous component. The latter is well represented by
a spheroidal halo with a constant density in the central regions and $r^{-2}$
profile in the outermost ones. The importance of the dark component
for the mass structure is found to be inversally proportional to the
galaxy luminosity but not directly dependent on its morphology. Similarly, the
inferred parameters of the DM distribution (i.e.: central density and core
radius) as well as those of the visible matter distribution (i.e.: the
{\it stellar} mass-to-light ratio)
are found
to be dependent only on the luminosity of the object considered. (Persic \&
Salucci
1990, 1992, 1993).

This systematic luminosity variation of the DM content
in galaxies can be used to assess with better accuracy the long debated issue
of the galaxy contribution to $\Omega_0$,
in that we are now able to estimate with great precision the mass of a
galaxy inside any fraction $0.2 \leq f \leq 2$ of its optical size. However,
to infer the total mass of a galaxy one  needs to know the
extent of its halo. From an observational point of view,  the dark halos are
gravitationally detected from the non-keplerian
profiles of rotation curves out to a couple of optical radii, at larger radii
the kinematics of galaxy pairs imply halos
extended  to hundreds of kpc, which turns out to be
the scale of their virialization radii. Therefore, in this paper
we will use  a strict upper limit to the
halo mass obtained by extrapolating the known mass
distributions out  to the virialization radii. Then, we will
integrate this (upper limit to the)
mass of galaxies over the corresponding luminosity function (hereafter LF), and
obtain therefore an estimate of $\Omega_0^{gal}$,
 the (maximum) galaxy contribution to the mean
density of the Universe. (Hereafter, since the estimate
of $\Omega_0^{gal}$ made in this paper
is independent of $h$, we just  assume a particular value
for the Hubble contant: $h=0.75$, in order to drop
  drop  for simplicity
the  explicit $h$ dependency of masses, densities, luminosities etc.)

\section{The mass structure of galaxies}

Remarkable insight into halo properties can be obtained by decomposing rotation
curves into disc (plus bulge) and halo contributions. This leads to estimating
the halo mass,
$M_h^{opt}$, within the optical radius ($R_{opt}$) and the core radius $a$.
Proper dynamical decompositions of
rotation curves, applied to samples of galaxies spanning large ranges of
luminosity, have yielded that DM increasingly dominates over luminous matter
within $R_{opt}$ for decreasing galaxy luminosity, obeying a scaling relation
of the sort of $M_h^{opt}/M_{lum} \propto L_B^{-x}$ with $x \simeq 0.6-0.7$
(Persic \& Salucci  1990). This implies different variations with
luminosity for the visible mass and for the dark mass. In particular, for DM it
has been found $M_h^{opt} \propto L_B^k$ with $k \simeq 0.6$ (see Ashman et al.
1993). This luminosity effect in the content of DM, found at $R_{opt}$, is able
to explain the outer shapes of rotation curves (Salucci \& Frenk 1989;
Casertano \& van Gorkom 1991). Direct and statistical mass modelling
of spiral rotation curves have shown that the DM halo is well
represented by the following density law:
$$
\rho(r)~ = ~\rho_0 ~ {a^2 \over a^2\,e^{-(r/a)^2} + r^2}
\eqno(1a)
$$
$$
\rho_0 ~= ~ 0.01 ~\biggl( {L_B\over L_B^*} \biggr)^{-0.9}
\eqno(1b)
$$
$$
a ~\simeq ~ 15 ~ (L_B/L_B^*)^{0.5}~~~{\rm kpc}.
\eqno(1c)
$$
with $log ~ L_*/L_\odot=10.77$. Recently is has been suggested that the DM
properties of
ellipticals are closely similar to those of spirals of comparable luminosity
(see Bertola et al. 1993). Therefore, at least to a first
approximation, galaxy halos can be considered as a belonging to a one-parameter
sequence, the luminosity (or alternatively the visible mass) being the ordering
parameter. For our purposes here we shall assume that, irrespective of their
 optical morphology, galaxies of the same luminosity have the same halos.

Although on the scale of the separation radii of galaxy pairs
(typically, several hundreds of kpc) the dynamics is totally dominated by the
DM (Charlton \& Salpeter 1991), no velocity turnoff, signalling the halo cutoff
radius, has yet been observed for isolated galaxies. Therefore no total halo
mass has been measured. To estimate an upper limit to the halo mass and
a  cutoff radius, we resort here
to the standard  model of collapse of  density fluctuations. As it is well
known in an
otherwise unperturbed FRW universe, an isolated spherical density
fluctuation  expands out to the turnaround radius $R_t$ and then collapse
back, reaching virialization after a few bounces at a radius $R_v=1/2 R_t$,
when the associated overdensity is about 150, for objects formed
at $z_f\simeq 1$. (Notice that, since $ R_v \propto (1+z_f)^{-1.5}$, if
the bulk of galaxy formation occurs at earlier redshifts, our procedure will
{\it overestimate}
all galaxy masses).

 At the farthest radii with measured rotation curves $\simeq 2 R_{opt}$, halos
have already attained their  asymptotic velocity, therefore we can infer $R_v$
from eq. (1) by means of the condition $<\rho>_{R_v}=150 ~ \rho_c ~ \Omega_0$,
with
$\rho_c $ being the critical density, and $\Omega_0=1$ for
the purpose of this paper. In detail, we
smooth the halo density field described by eqs. (1) over a gaussian window of
radius $R_v$ such that the mean density within the window corresponds to 150
times the cosmological mean.
For a sample of about 60 galaxies with excellent surface photometry and
rotation curve data (Persic \& Salucci 1994) the resulting values are in the
range 60-200 kpc for luminosities in the range $8.5 \leq {\rm log}L_B/L_B^*
\leq 11.5$. These radii are compatible with the separation radii in galaxy
pairs, but are inconsistent with the prejudice of
 halo radii of size of about 1  Mpc often present in the literature.
The halo masses
encompassed by our estimated virial radii are in the range $11 \leq {\rm log}
M_h/M_\odot \leq 12.5$. This mass range is much more compressed than the
corresponding range in luminosity, implying $M_h \sim 10^{12} (L_B/L_B^*)^{0.5}
M_\odot$. This mass-luminosity relation in turn implies that the halo $M/L$
ratio is not constant but depends on luminosity according to $M_h/L_B ~\simeq ~
100~ (L_B/L_B^*)^{-0.5}$ (in solar units). This means, of course, that for
galaxies the mass function and the LF are not parallel to each other, in
particular reconciling the mass function with the scenario of an hierarchical
galaxy formation (Ashman et al. 1993). Of course,
environmental effects (e.g.,
interactions) may have tidally removed the outskirts of halos
that now can be less extended than $R_v$, however, this dynamical
evolution is irrelevant for the scope of this paper.

\section{The contribution of galactic halos to $\Omega_0$}

The contribution of virialized galactic halos to $\Omega_0$ can now be
estimated directly by evaluating the integral
$$
\Omega_0^{gal}(L) = {1\over \rho_c}
\int_{L_{min}}^{L} \phi(L) \biggl({ M_h(L)\over L} \biggr) ~L~dL\,,
$$
where $\phi(L)$ is the standard galaxy LF, which we take from Efstathiou et al.
(1988). In Figs (1) and (2)  we show, as a function of log$L$ and
log $R_v$, the integral contribution to
$\Omega_0$ from galaxies of luminosity $\leq L$
and an halo size   $\leq R_v$. It is obvious
that $\Omega_0^{gal}(L)$ quickly saturates at 0.1 for $L> L_*$
(or consequently $R_v > 0.1 ~ Mpc $). This estimate firmly
suggests a low  contribution to the cosmological density coming  from
normal galaxies and that the dark halos sizes are much smaller
than the galaxy clustering lenght (about 7 Mpc).
In conclusion, the matter contained in visible
galaxies is unable to close the Universe. One should invoke
 the existence of a
significant new population of low-$L$, low-surface-brightness galaxies with a
steep luminosity function (e.g., Phillips 1994) in order to  raise
substantially our estimate.

There is a separate aspect of interest to our result. Our deduced
 value of $\Omega_0^{gal} \leq 0.1$ is comparable with that
of the standard Big-Bang
nucleosynthesis: $\Omega_{bar} = 0.06\,h_{50}^{-2}$, so that
the DM in galactic halos needs not to be of exotic nature. This, in
connection with the very small fraction of visible baryons in identified
structures ($\Omega_* \simeq 0.003$, Persic \& Salucci 1992) and with the
repeatedly negative Gunn-Peterson tests, suggests that the missing baryons may
be locked in the dark halos of galaxies, leading to an attractive
solution of the `missing baryons' problem.

\end{document}